# Parametric Synthesis of Computational Circuits for Complex Quantum Algorithms

**Cesar Borisovich Pronin**

**Andrey Vladimirovich Ostroukh**

MOSCOW AUTOMOBILE AND ROAD CONSTRUCTION STATE TECHNICAL UNIVERSITY (MADI)., 64, Leningradsky prospect, Moscow, Russia

**Abstract:** At the moment, quantum circuits are created mainly by manually placing logic elements on lines that symbolize quantum bits.

The purpose of creating Quantum Circuit Synthesizer "Naginata" was due to the fact that even with a slight increase in the number of operations in a quantum algorithm, leads to the significant increase in size of the corresponding quantum circuit. This causes serious difficulties both in creating and debugging these quantum circuits.

The purpose of our quantum synthesizer is enabling users an opportunity to implement quantum algorithms using higher-level commands. This is achieved by creating generic blocks for frequently used operations such as: the adder, multiplier, digital comparator (comparison operator), etc. Thus, the user could implement a quantum algorithm by using these generic blocks, and the quantum synthesizer would create a suitable circuit for this algorithm, in a format that is supported by the chosen quantum computation environment. This approach greatly simplifies the processes of development and debugging a quantum algorithm.

The proposed approach for implementing quantum algorithms has a potential application in the field of machine learning, in this regard, we provided an example of creating a circuit for training a simple neural network.

Neural networks have a significant impact on the technological development of the transport and road complex, and there is a potential for improving the reliability and efficiency of their learning process by utilizing quantum computation, through the introduction of quantum computing.

**Keywords:** quantum synthesizer, quantum circuit synthesis, quantum machine learning, quantum algorithms, neural networks, quantum computing, Grover's algorithm.

**Introduction**

The goal of project "Naginata - Quantum Circuit Synthesizer" [1] is to create a prototype system for synthesizing complex quantum circuits. One of the applications of this system will be in parametric synthesis of quantum circuits for finding the weights of a neural network (perceptron) of a given topology. Quantum circuits synthesized by the program are exported to a *.qasm text file in the form of OPENQASM 2.0 code, compatible with IBM Quantum - a cloud quantum computing environment.

To create quantum circuits that implement complex algorithms, a large number of available quantum bits is required for building the circuit. In this regard, the IBM Quantum cloud service was chosen as a platform for debugging [2]. IBM Quantum allows users to perform quantum computing tasks on state-of-the-art quantum computer and quantum simulator prototypes.

Since all current generation quantum computing environments are significantly limited by the number of available quantum bits, some circuit optimization is often necessary to demonstrate complex circuits. One of the methods for optimizing circuits is to use quantum registers with the minimum required size. Thus, the synthesized circuits of typical blocks included in a quantum algorithm must change, based on the size of these registers.

This paper demonstrates the circuits, which lay in the basis of algorithms, created to scale these circuits, depending on the input parameters. For clarity, these circuits were implemented in the "Quirk" quantum simulator, which also supports a certain set of typical operations.

As an example, these synthesis methods were used in an algorithm to find weights suitable for training the neural network on **Fig. 7**.

**Adder circuit**

IBM Quantum Composer currently does not have built-in methods for performing arithmetic operations, therefore they need to be created. It is necessary to form scalable circuits for the adder, multiplier, and comparison operator. These operations must support the generation of their reversed forms - operations with a reverse sequence of logic gates. This is required for utilizing the reversibility principle in synthesized quantum circuits. As an example, reversed circuits are necessary for completing the oracle function in our implementations of Grover's algorithm [3].

The adder circuit, which does not require auxiliary qubits (ancilla) was taken from the built-in example in the Quirk quantum simulator [4].

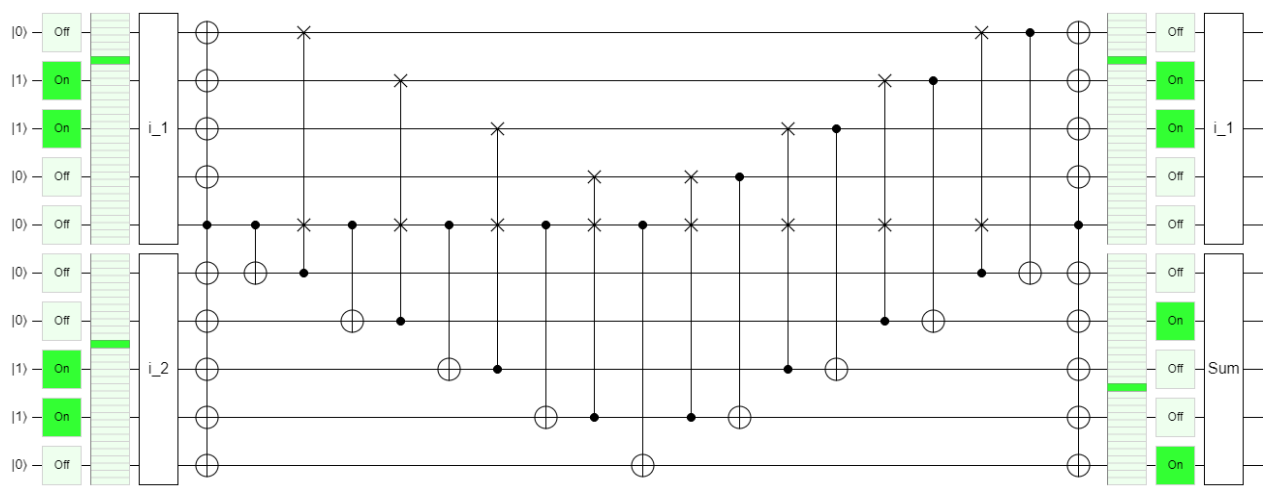

*Fig. 1. Adder circuit - calculates the sum of registers i_1 and i_2, and writes the sum to i_2, leaving i_1 unchanged*

In the adder circuit, for compatibility with IBM Composer, it is necessary to decompose the controlled Pauli-X gates on several target qubits into a set of sequential operations (**Fig. 2**). This is done by the "multi_target_gate" function:

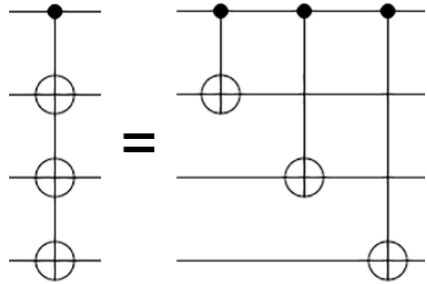

*Fig. 2. Transformation done by the "multi_target_gate" function*

**Multiplier circuit**

In the process of calculations, it often becomes necessary to use auxiliary qubits, known as ancilla, for intermediate calculations, the values of ancilla qubits are later cleared by utilizing the reversibility principle. So, in order to reduce the number of quantum bits used, we place ancilla qubits in a separate register, which is shared by sequential operations. When the circuit is built, this register is added to the circuit using the "plot_ancilla()" method applied to the exported object of the "Circuit" class. In our synthesizer, ancilla qubits are declared by assigning the value "True" to their "is_ancilla" property.

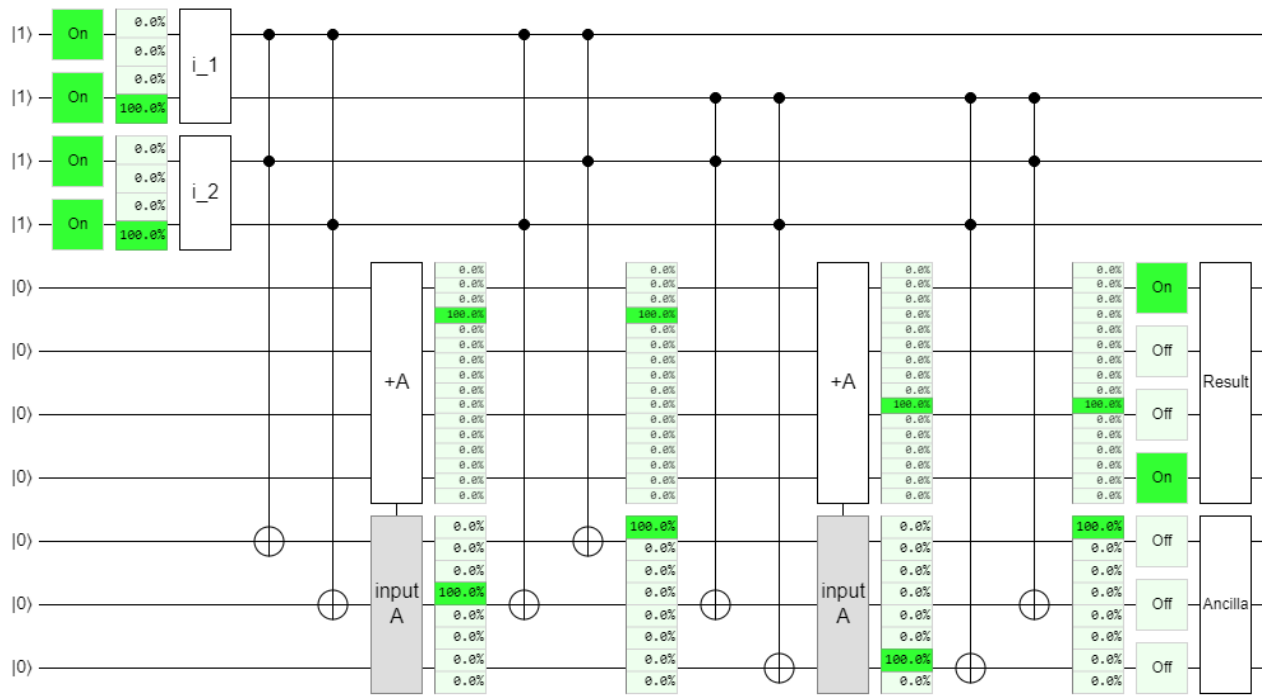

*Fig. 3. Multiplier circuit – calculates the product of registers i_1 and i_2 in register "Result"*

The multiplication algorithm is based on the principle of binary long multiplication, combined with the adder from **Fig. 1**. Qubit multiplication is performed using Toffoli gates, the product is written to ancilla qubits, then the adder transfers ancilla values to the resulting product register, after which, using the reversibility principle, the ancilla register is cleared for later use. The quantum synthesizer implements parametric QASM code generation depending on the input register qubits and their size. The "multiplier_asymmetric" function generates circuits for multiplying different sized registers.

### Digital comparator circuit

To compare values of different registers, the circuit of a classical digital comparator [5-6] was interpreted as a quantum circuit ("if_equal" function) (**Fig. 4**).

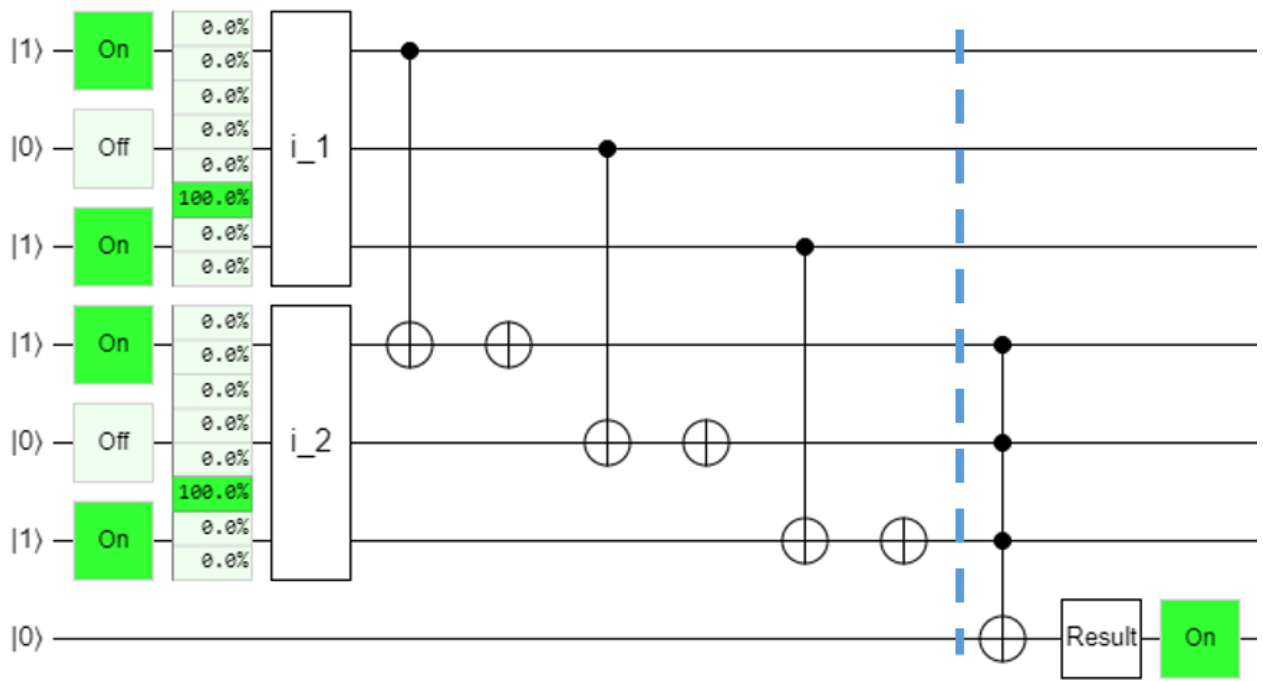

*Fig. 4. Quantum circuit for comparing qubits of two 3-qubit registers i_1 and i_2, writes the result in register i_2 (before the dotted line)*

# Multiple-Control Toffoli gate circuit

To change the sign of the target states in the quantum register allocated for Grover's algorithm, a Pauli-Z gate controlled by a set of qubits is needed. It can be obtained by adding a Hadamard gate on each side of a Toffoli gate's target element (Pauli-X).

Usually, to create an element with a large number of control qubits, a set of Toffoli gates with ancilla qubits as a buffer is used (**Fig. 5**). Our "multi_control_gate_3cx" circuits are generated similar to this concept, but they use a 4-qubit Toffoli implementation (**Fig. 6**) to reduce the number of required ancilla qubits.

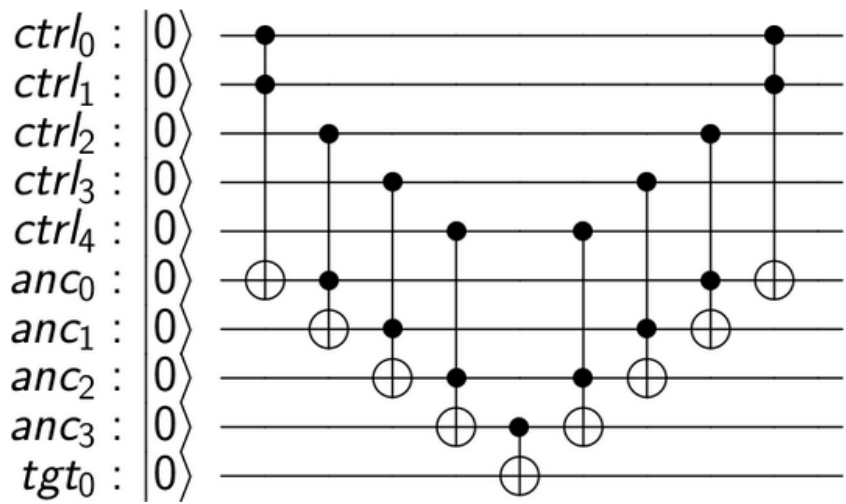

*Fig. 5. A general multiple-control Toffoli gate circuit [7]*

The standard Toffoli gate has 2 control qubits. There is a Toffoli implementation (**Fig. 6**) with three control qubits. This implementation is obtained by visualizing the "mct" function for 4 qubits from the "Qiskit" framework [8-9]. In "Naginata", it is created using the "Toffoli_4q" function, or by adding a "GName.CCCX" logic gate to a circuit.

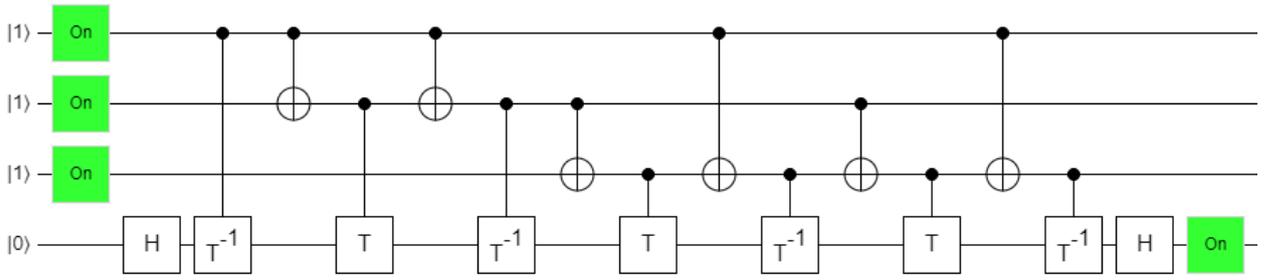

*Fig. 6. 4-qubit Toffoli circuit*

**Example use case scenario of the described functions in development of a quantum algorithm (on the example of a perceptron training circuit)**

Now, parametric versions of the described circuits can be used as building blocks for building more complex algorithms. Let's look at an example of finding weights in a statically defined neural network with the topology described on **Fig. 7**. It is implemented in the file "qnn_static_test.py".

The goal of training this neural network is to find the coefficients $w_i$ that satisfy inequality (1):

$$(I_1 w_1 + I_2 w_2) * w_3 \geq Ac \qquad (1)$$

To simplify the analysis of the algorithm's results, we implement only condition (2) [10-12]:

$$(I_1 w_1 + I_2 w_2) * w_3 = Ac \qquad (2)$$

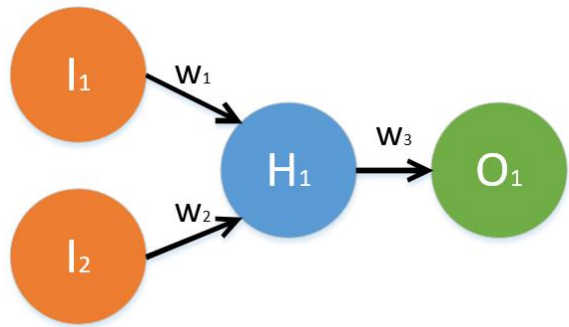

*Fig. 7. Topology of the example neural network*

At the input of the algorithm, the values are: $I_1 = 11_2 = 3_{10}$; $I_2 = 10_2 = 2_{10}$; $Ac = 000110_2 = 6_{10}$

At the output of the algorithm, the following values were obtained as solutions: 010010, 011100, 100001, 110100. They correspond to peaks in the diagram on **Fig. 8**. Based on the order in which the registers are defined for this circuit in file "qnn_static_test.py" and the bit numbering order in the IBM Quantum environment, the obtained values should be divided into equal 2-bit registers corresponding to desired values of $w_1, w_2$ и $w_3$, in the manner indicated in **table 1**.

Table 1. The order of splitting the measured bit strings into w values

| measured | $w_3$ | $w_2$ | $w_1$ |
|---|---|---|---|
| 010010 | 01 | 00 | 10 |
| 011100 | 01 | 11 | 00 |
| 100001 | 10 | 00 | 01 |
| 110100 | 11 | 01 | 00 |

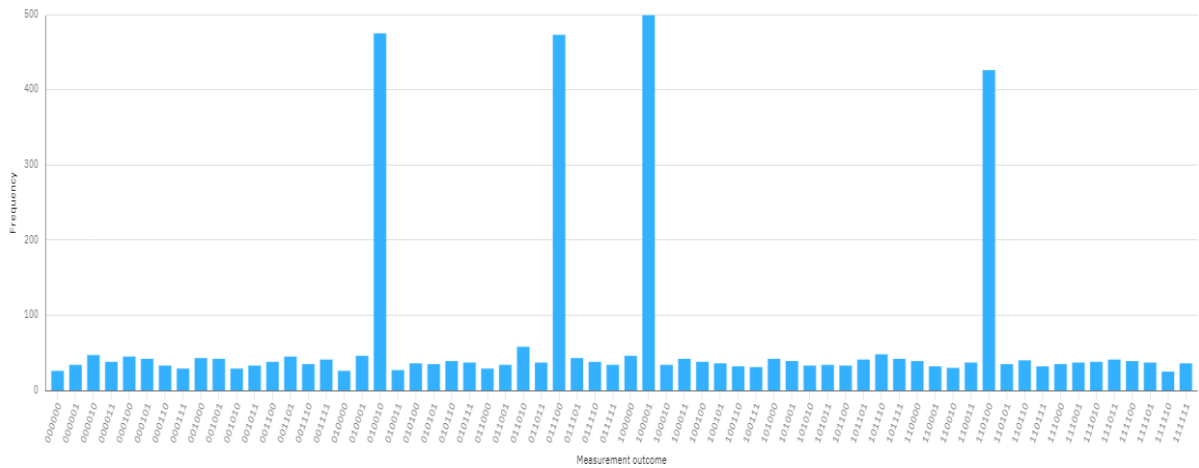

*Fig. 8. The spread of the algorithm's measured values after 4096 iterations (shots = 4096)*

**Conclusions**

Demonstrated quantum circuits, lay in the basis of algorithms for scaling these circuits, depending on their input parameters. These algorithms were used in our quantum synthesizer prototype, called "Naginata". For clarity, these circuits were implemented in the Quirk quantum simulator, which also supports a certain set of typical operations.

As a possible application example, of the developed synthesizer, the presented methods were used in an algorithm that finds weights suitable for training an example neural network (perceptron) on **Fig. 7**. The distribution diagram of the output values, which was obtained after running the synthesized circuit on IBM's 100-qubit quantum simulator "simulator_mps" is provided on **Fig. 8**.

The prototype of the quantum synthesizer was developed with the possibility of its further modification and expansion of capabilities by adding new functions that implement typical operation blocks.

Among the ways of further developing this project, it is important to note the development of more advanced versions of the described typical blocks that require a minimal number of operations and ancilla qubits, as well as the addition of new blocks to expand the range of applications of the quantum synthesizer. This would make it possible to form necessary tools for the transition from creating quantum circuits at the level of logic gates to quantum programming using pre-prepared compositions of these logic gates.

The use of computational advantages of quantum computers in machine learning could significantly optimize artificial intelligence models and improve the accuracy of their training, which in turn would make these models more reliable and versatile. In particular, this can increase the speed and accuracy of object recognition systems that are actively being introduced into unmanned vehicles and driver assistance systems, as well as many other promising developments in the transport and road complex and other spheres of economy.

**References**

[1] C.B. Pronin: Naginata - Quantum Circuit Synthesizer. GitHub, URL: https://github.com/MexIvanov/Naginata-QCS

[2] IBM Quantum Computing, URL: https://www.ibm.com/quantum

[3] Grover L.K.: A fast quantum mechanical algorithm for database search. Cornell University Library. URL: https://arxiv.org/abs/quant-ph/9605043

[4] Quirk – online quantum computer simulator. URL: http://algassert.com/quirk

[5] Binary Comparators. Learnabout electronics. URL: https://learnabout-electronics.org/Digital/dig43.php

[6] Digital Comparator. Electronics Tutorials. URL: https://www.electronics-tutorials.ws/combination/comb_8.html

[7] M. Nielsen, I. Chuang.: Quantum Computation and Quantum Information: 10th Anniversary Edition. Cambridge University Press, 2010, p. 177 – 4.3 Controlled operations. ISBN 978-1-107-00217-3 Hardback.

[8] Qiskit - open-source SDK for working with quantum computers at the level of pulses, circuits, and application modules. URL: https://qiskit.org/documentation/

[9] Quantum Computing Stack Exchange. URL: https://quantumcomputing.stackexchange.com/questions/2177/how-can-i-implement-an-n-bit-toffoli-gate

[10] C. B. Pronin, O. I. Maksimychev, A. V. Ostroukh, A. V. Volosova and E. N. Matukhina, "Creating Quantum Circuits for Training Perceptron Neural Networks on the Principles of Grover's Algorithm," 2022 Systems of Signals Generating and Processing in the Field of on Board Communications, 2022, pp. 1-5, doi: 10.1109/IEEECONF53456.2022.9744279.

[11] C.B. Pronin, A.V. Ostroukh: Development of quantum circuits for machine learning of neural networks with perceptron architecture based on the principles of Grover's algorithm. Industrial Automatic Control Systems and Controllers. (in Russian) 2021, No. 12, P. 3-9. URL: http://asu.tgizd.ru/

[12] C.B. Pronin, A.V. Ostroukh: Creation and training of quantum neural networks based on Grover's algorithm. Industrial Automatic Control Systems and Controllers. (in Russian) 2021, No. 11, P. 39-43. URL: http://asu.tgizd.ru/

**Author details**

**Andrey Vladimirovich Ostroukh**, Russian Federation, full member RAE, Doctor of Technical Sciences, Professor, Department «Automated Control Systems». State Technical University – MADI, 125319, Russian Federation, Moscow, Leningradsky prospekt, 64. Tel.: +7 (499) 151-64-12. http://www.madi.ru , email: ostroukh@mail.ru , ORCID: https://orcid.org/0000-0002-8887-6132

**Cesar Borisovich Pronin**, Russian Federation, PhD student, Department «Automated Control Systems». State Technical University – MADI, 125319, Russian Federation, Moscow, Leningradsky prospekt, 64. Tel.: +7 (499) 151-64-12. http://www.madi.ru , email: caesarpr12@gmail.com , ORCID: https://orcid.org/0000-0002-9994-1032